\documentclass{article}

\usepackage[preprint]{neurips_2024}

\usepackage[utf8]{inputenc} 
\usepackage[T1]{fontenc}    
\usepackage{hyperref}       
\usepackage{url}            
\usepackage{booktabs}       
\usepackage{amsfonts}       
\usepackage{nicefrac}       
\usepackage{microtype}      
\usepackage{xcolor}         
\usepackage{subfig}
\usepackage{graphicx}

\title{Double-Condensing Attention Condenser: Leveraging Attention in Deep Learning to Detect Skin Cancer from Skin Lesion Images}

\author{%
  Chi-en Amy Tai*\\
  Department of Systems Design Engineering\\
  University of Waterloo\\
  Waterloo, ON\\
  \texttt{amy.tai@uwaterloo.ca} \\
  \And
  Elizabeth Janes*\\
  Department of Systems Design Engineering\\
  University of Waterloo \\
  \And
  Chris Czarnecki\\
  Department of Systems Design Engineering\\
  University of Waterloo \\
  \And
  Alexander Wong\\
  Department of Systems Design Engineering\\
  University of Waterloo \\
  \And
  \textsuperscript{*} All authors contributed equally.\\
}

\begin{document}

\maketitle

\begin{abstract}
Skin cancer is the most common type of cancer in the United States and is estimated to affect one in five Americans. Recent advances have demonstrated strong performance on skin cancer detection, as exemplified by state of the art performance in the SIIM-ISIC Melanoma Classification Challenge; however these solutions leverage ensembles of complex deep neural architectures requiring immense storage and compute costs, and therefore may not be tractable. A recent movement for TinyML applications is integrating Double-Condensing Attention Condensers (DC-AC) into a self-attention neural network backbone architecture to allow for faster and more efficient computation. This paper explores leveraging an efficient self-attention structure to detect skin cancer in skin lesion images and introduces a deep neural network design with DC-AC customized for skin cancer detection from skin lesion images. The final model is publicly available as a part of a global open-source initiative dedicated to accelerating advancement in machine learning to aid clinicians in the fight against cancer. Future work of this research includes iterating on the design of the selected network architecture and refining the approach to generalize to other forms of cancer.
\end{abstract}

\section{Introduction}
Skin cancer is the most common type of cancer in the United States and is estimated to affect one in five Americans~\cite{SkinCancer}. In the United States alone, the annual cost for treating skin cancer is estimated at \$8.1 billion~\cite{SkinCancer}. The main types of skin cancer are squamous cell carcinoma, basal cell carcinoma, and melanoma~\cite{SkinCancerNIHStats}. Although melanoma is less common than the other skin cancer types, melanoma causes most skin cancer deaths due to its invasive nature~\cite{SkinCancerNIHStats}. However, the chances of surviving five years after a melanoma skin cancer diagnosis significantly improve if the cancer is detected at an early stage~\cite{SkinCancerNIHStats}. At later stages, the cancer may have spread to nearby skin and organs leading to more severe consequences for the patient and more expensive and complex treatments~\cite{SkinCancerNIHStats}. On the other hand, a standard surgery can be used to remove the lesion for early-stage melanoma~\cite{SkinCancerNIHStats}. 

\begin{figure}
 \begin{center}
 \subfloat[\centering Melanoma skin cancer]{{\includegraphics[width=.4\linewidth]{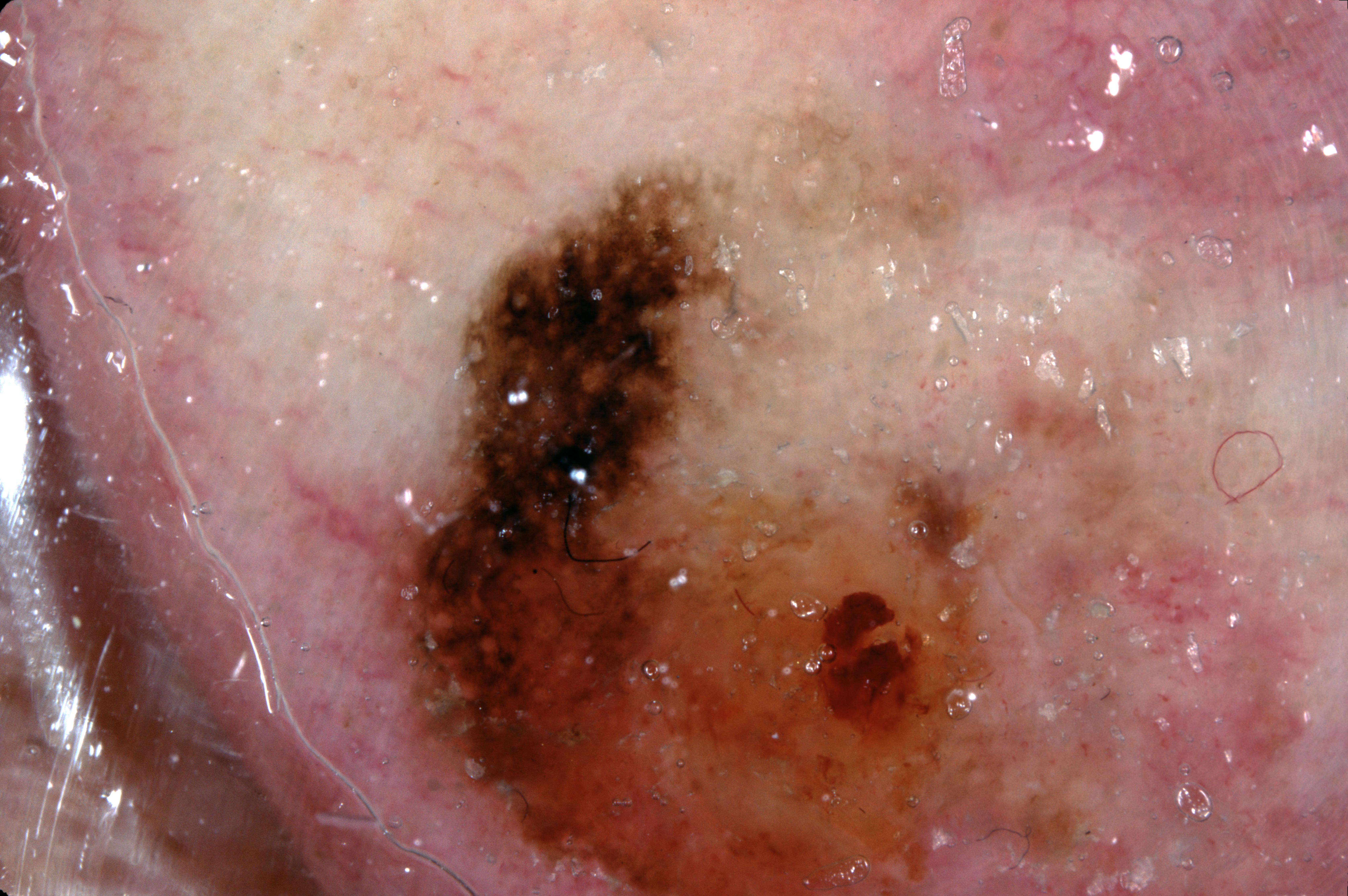} }}%
 \qquad
 \subfloat[\centering Non skin cancer]{{\includegraphics[width=.4\linewidth]{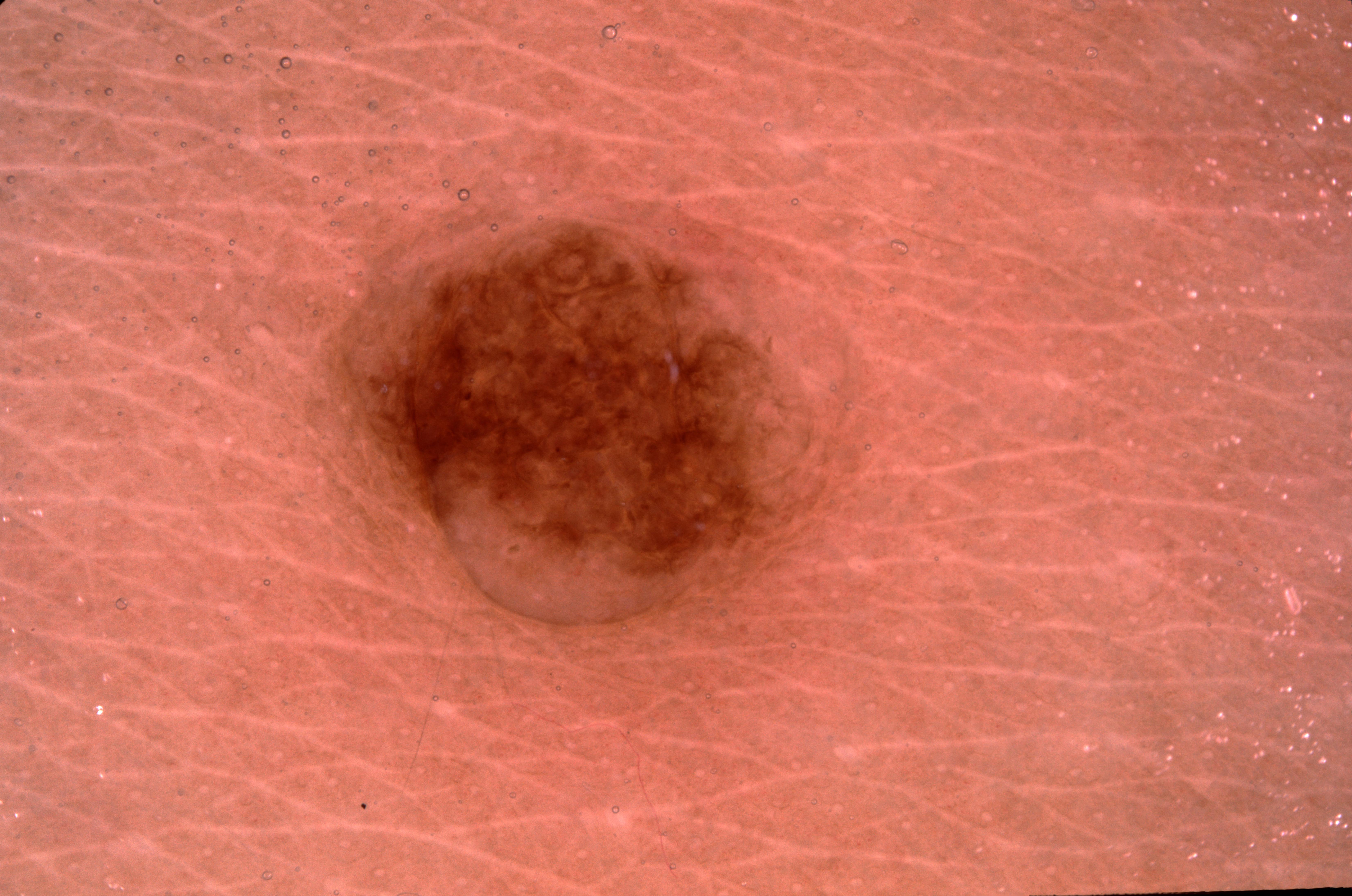} }}%
 \end{center}
 \caption{Example skin lesion images.}
 \label{fig:sample-skin-images}
\end{figure}

The current method to diagnose skin cancer relies on a dermatologist looking at a tissue sample of a skin lesion under a microscope. An example of a melanoma skin cancer and non skin cancer image is shown in Figure~\ref{fig:sample-skin-images}. Unfortunately, the current diagnostic process requires a skin biopsy where suspect skin cells are removed from the patient which can lead to anxiety and physical discomfort. Furthermore, the sole dependence on the human dermatologist results in potential erroneous diagnosis as there is potential biases and high uncertainty in clinical judgement~\cite{human-judgement-problem-skin-cancer-2,human-judgement-problem}. Subsequently, there has been a rise in research focused on computer-assisted diagnosis of melanoma skin cancer~\cite{sc-research-computer-assisted}. With the introduction of ensemble network architecture designs, a large shift in the high-performing skin cancer detection research has been centered around large computationally expensive ensemble designs. Realistically, these approaches may not be tractable for clinical deployment due to the high storage requirements which preclude integration of ensemble models into mobile devices and portable dermoscopy devices. Likewise, computational complexity of ensemble models increases diagnostic latency which may also limit their clinical utility.

Image classification tasks require non-linear relationships to be captured. Spatial relationships cannot be effectively represented using linear models~\cite{lecunDeepLearning2015}, therefore deep learning is preferred for image classification tasks~\cite{krizhevskyImageNetClassificationDeep2012}. A substantial problem which affects all neural network architectures is their tendency to overfit to the training data~\cite{goodfellow2016deep}. Despite this limitation, it has been shown that for many complex tasks with non-linear non-convex loss landscapes, most neural network architectures (including the  transformer model) tend to benefit from a larger number of parameters, as increasing the depth of the network decreases the inductive bias imposed onto the model~\cite{kaplanScalingLawsNeural2020}. Due to the bias-variance tradeoff, it is also desired for such models to be trained on substantially large datasets to avoid overfitting~\cite{goodfellow2016deep}. Thereby, these criteria results in architectures of significant depth trained on large datasets, necessitating availability of significant compute resources at training time. The large  number of parameters in such architectures also results in significant storage requirements and demands significant compute resources during inference. Both of these requirements impose constraints on the devices where high performance neural networks for image classification can be deployed. This is particularly relevant in the context of resource-constrained environments and mobile devices which may be used for clinical skin cancer detection, such as portable dermoscopy devices. 

The TinyML space specifically examines machine learning models in resource-constrained environments and a recent innovation in the TinyML deep learning space is the Double-Condensing Attention Condenser (DC-AC) architecture design, shown to achieve high inference throughput despite its small size~\cite{wongFasterAttentionWhat2022}. The design of the architecture was based on machine-driven exploration and leverages a DC-AC mechanism~\cite{wongFasterAttentionWhat2022}. This mechanism introduces another Attention Condenser module and calculates selective attention for the outputs of both condenser modules which leads to highly condensed feature embeddings~\cite{wongFasterAttentionWhat2022}. Learning more condensed feature embeddings improves the representational performance and efficiency of the network, which is especially useful for skin cancer as it improves attention on key characteristics associated with cancer while providing a wide field of vision. When compared to other state of the art efficient networks such as MobileOne-S1, this architecture was also shown to outperform on the top-1 accuracy for ImageNet with a higher inference throughput and smaller network architecture size~\cite{wongFasterAttentionWhat2022}.

Motivated by the DC-AC architecture, this paper explores leveraging the efficient self-attention structure to detect skin cancer in skin lesion images. We leverage the ISIC 2020 dataset~\cite{rotemberg_patient-centric_2021} to investigate the performance of this smaller architecture on the unseen ISIC 2020 test set. We find that it performs superior to both the previously released Cancer-Net SCa network architecture designs and MobileViT-S, and hence, we introduce the DC-AC, a deep neural network design customized for skin cancer detection from skin lesion images that is publicly available at \url{https://alexswong.github.io/Cancer-Net/}.

\section{Related Works}
State of the art performance in the Society for Imaging Informatics in Medicine (SIIM) - International Skin Imaging Collaboration (ISIC) Melanoma Classification Challenge exemplifies the recent advancements in skin cancer detection, however such deep learning solutions leverage ensemble network architecture designs, requiring immense storage and compute costs. One of the top performers in this challenge used an ensemble network architecture design of EfficientNet B3-B7, se\_resnext101, and resenest101, while emphasizing validation methods, model target selection, and the data preparation pipeline to find the optimal solution, which resulted in an Area Under the ROC Curve (AUROC) score of 0.9490~\cite{ha_identifying_2020}. Similarly, the second place winner achieved an AUROC score of 0.9485 by leveraging an ensemble network architecture design of EfficientNet-B6, EfficientNet-B7, and a model trained on pseudolabelled test data~\cite{pan_siim-isic_2020,pan_siim-osic_2023}. Despite the high performance scores, these approaches are computationally expensive given the high storage costs for large ensemble models and corresponding high inference time. 

On the other hand, in 2020, Lee et al. released a suite of three Cancer-Net SCa network architecture designs tailored for skin cancer detection and designed using generative synthesis~\cite{leeCancerNetSCaTailoredDeep2020}. Notably, the macroarchitecture designs of the three networks are drastically different with Cancer-Net SCa-A and Cancer-Net SCa-B leveraging the projection-expansion-projection-expansion (PEPE) design pattern and Cancer-Net SCa-C using a highly efficient self-attention architecture design with attention condensers~\cite{leeCancerNetSCaTailoredDeep2020}. As such, the three designs have different performance-efficiency tradeoffs; Cancer-Net SCa-C has the lowest computational complexity whereas Cancer-Net SCa-A has the highest sensitivity and NPV but Cancer-Net SCa-B has the highest accuracy and PPV and lowest architectural complexity~\cite{leeCancerNetSCaTailoredDeep2020}. The Cancer-Net SCa architecture designs were reported to achieve strong performance on skin cancer detection with performance of up to 92.8 on malignant sensitivity~\cite{leeCancerNetSCaTailoredDeep2020}. However, when applied to the completely unseen ISIC 2020 test set, the pretrained Cancer-Net SCa architecture designs only have an acceptable ability~\cite{mandrekarReceiverOperatingCharacteristic2010} to discriminate between skin cancer and non skin cancer images (AUROC=0.77).

\begin{figure}
 \begin{center}
 \subfloat[\centering Benign skin lesions that look malignant]{{\includegraphics[width=.42\linewidth]{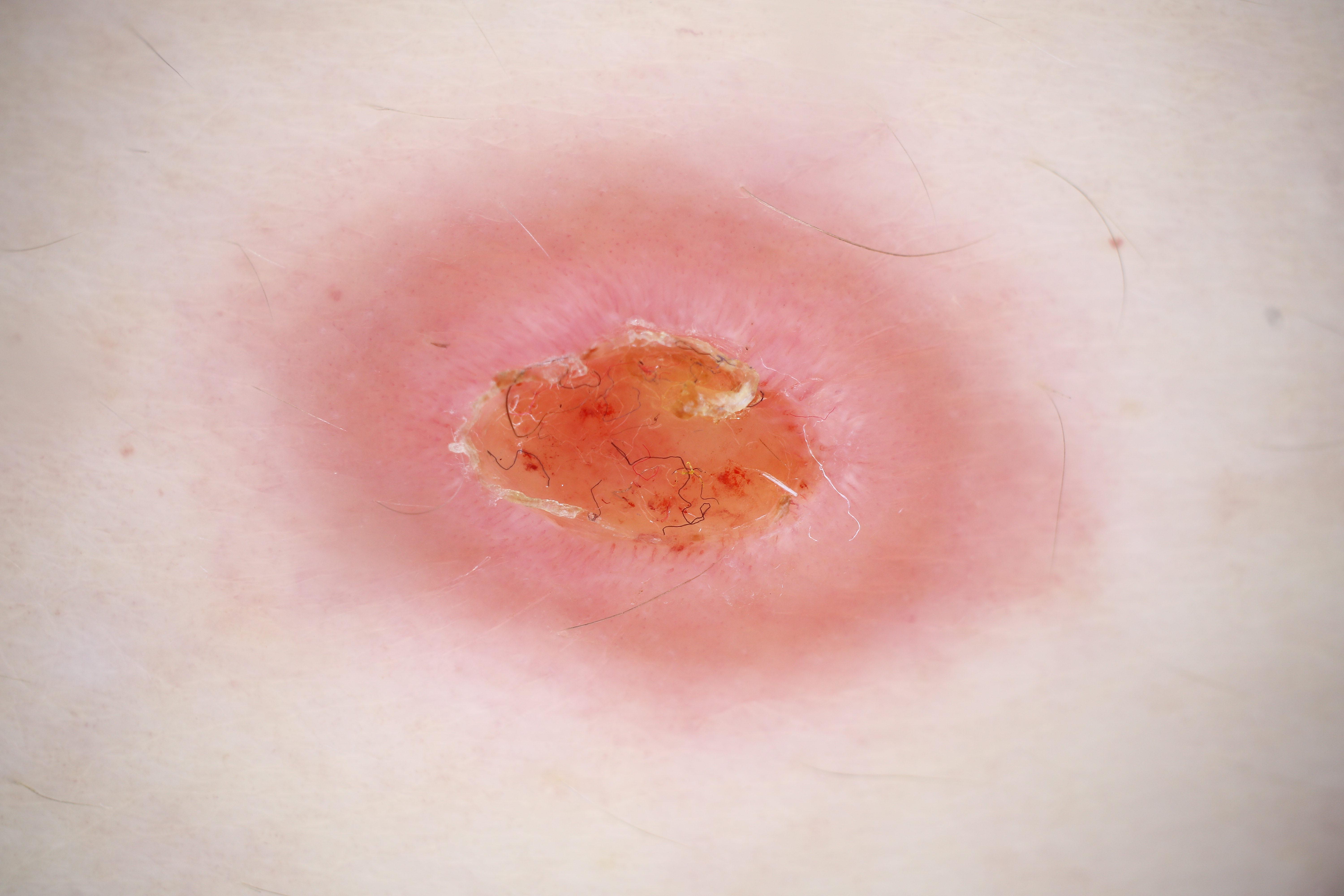}} {\includegraphics[width=.5\linewidth]{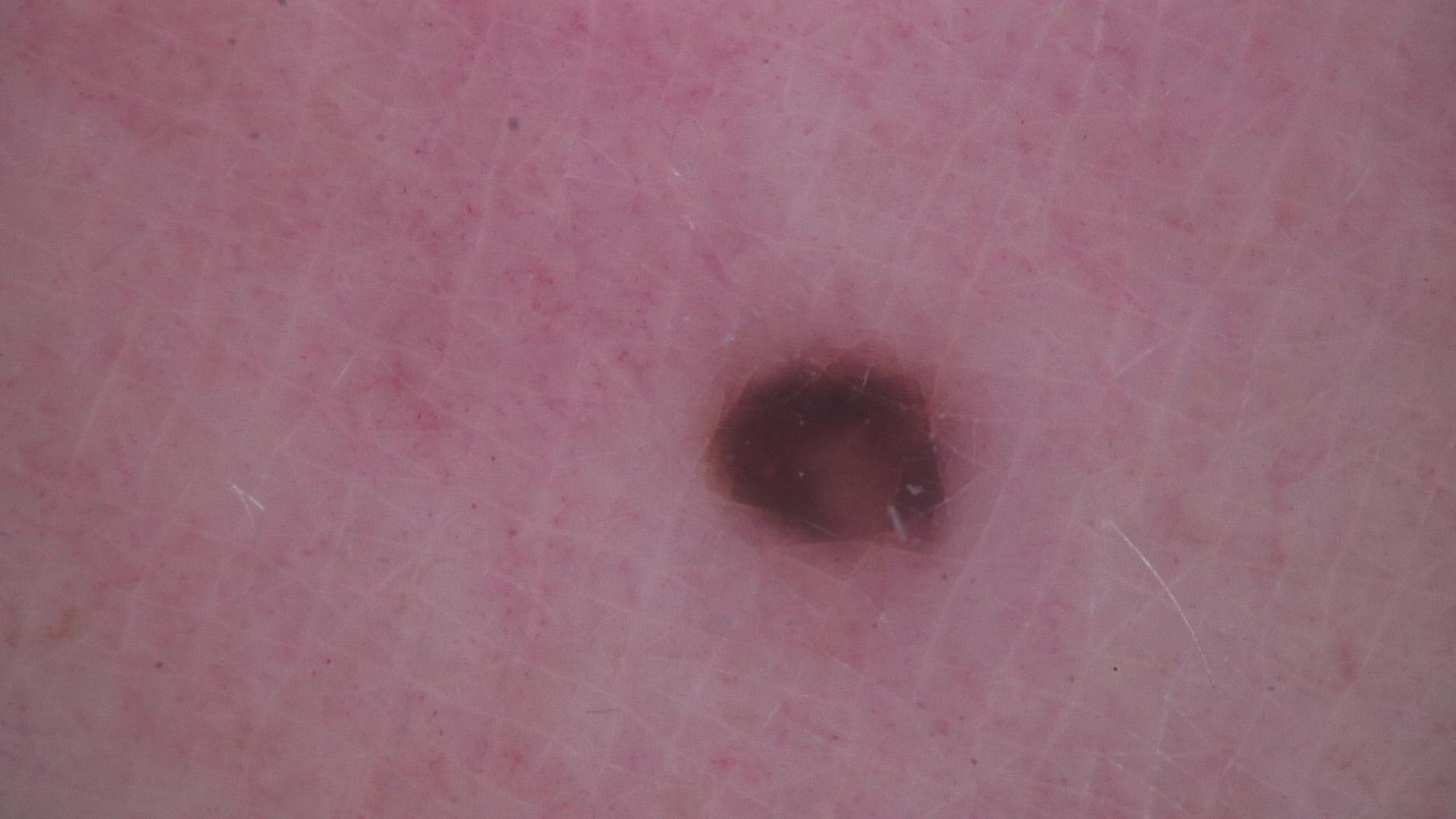}}}%
 \qquad
 \subfloat[\centering Malignant skin lesions that look benign]
 {{\includegraphics[width=.5\linewidth]{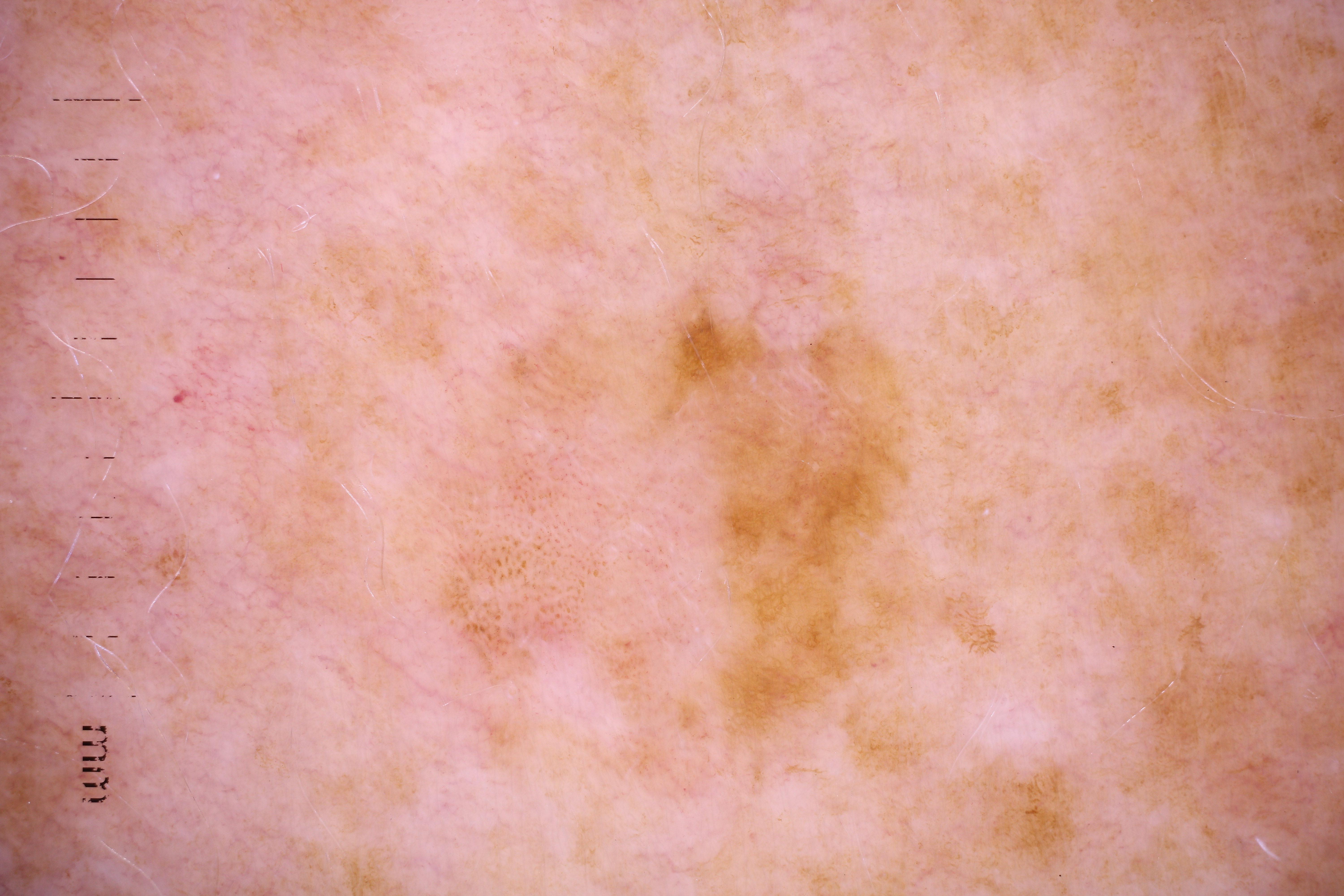}} {\includegraphics[width=.5\linewidth]{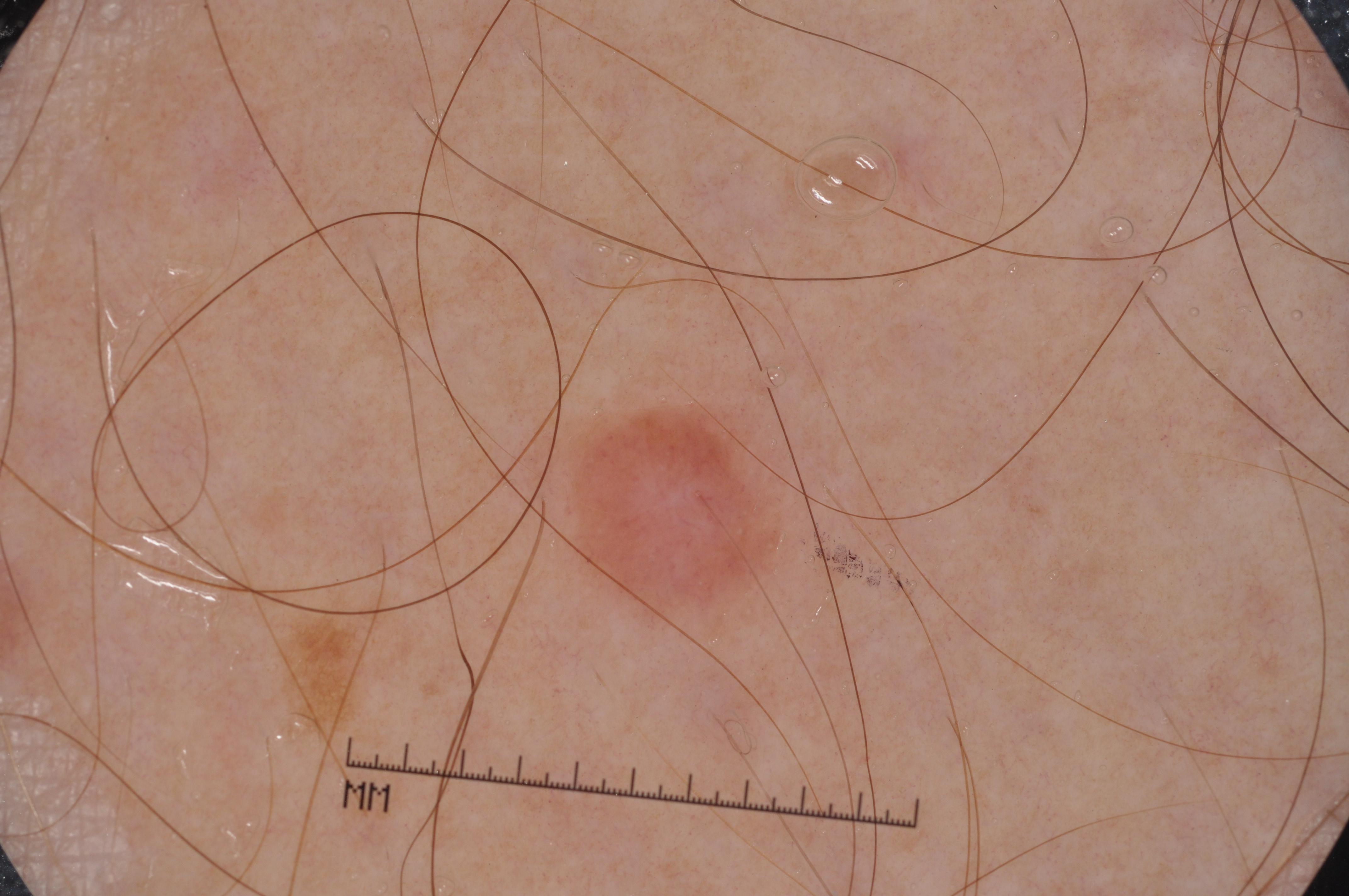}}}%
 \end{center}
 \caption{Sample skin lesions that look like the other class.}
 \label{fig:sample-confused-images}
\end{figure}

The DC-AC architecture design was recently introduced with the benefit of balancing network size and performance~\cite{wongFasterAttentionWhat2022}. This architecture design builds on the original idea of Attention Condensers. Attention Condensers include condenser layers, embedding layers, and expansion layers~\cite{wongTinySpeechAttentionCondensers2020}. The condenser layers project layer inputs into a lower-dimensional space with the preservation of activations in close proximity of other high activations~\cite{wongTinySpeechAttentionCondensers2020}. The embedding layer produces a condensed embedding, accounting for both local and cross-channel activation dependencies~\cite{wongTinySpeechAttentionCondensers2020}. The expansion layers project the produced embedding back to the original input space dimensions~\cite{wongTinySpeechAttentionCondensers2020}. An attention score (selective attention) is then calculated between the original inputs and the condenser module output projections~\cite{wongTinySpeechAttentionCondensers2020}. Unlike the original Attention Condensers, the DC-AC architecture introduces DC-AC modules which contain another Attention Condenser module and calculates selective attention for the outputs of both condenser modules~\cite{wongFasterAttentionWhat2022}. The condensers are not identical in the sense that one of them applies two embedding layers after projecting the inputs into the lower-dimensional space and leads to the learning of more condensed feature embeddings~\cite{wongFasterAttentionWhat2022}. By having a columnar network architecture, different branches in the early layers learn disentangled embeddings that are then merged in the deeper layers to gradually increase the channels. Henceforth wider areas of the original image are then covered by the self-attention blocks. This kind of architecture design is useful for skin cancer as the selective attention and condensation improves focus on key characteristics associated with cancer and also provides a wider field of vision in the image. This provides a significant advantage in the challenging task of detecting skin cancer, as identifying the class of a specific skin lesion is a difficult problem as seen in Figure~\ref{fig:sample-confused-images} where benign skin lesions can look malignant and malignant skin lesions can look benign.

\section{Materials and Methods}
\subsection{Data Preparation}
\begin{figure}
 \begin{center}
 \subfloat[\centering Images with hair obscuring parts of the skin lesion]{{\includegraphics[width=.5\linewidth]{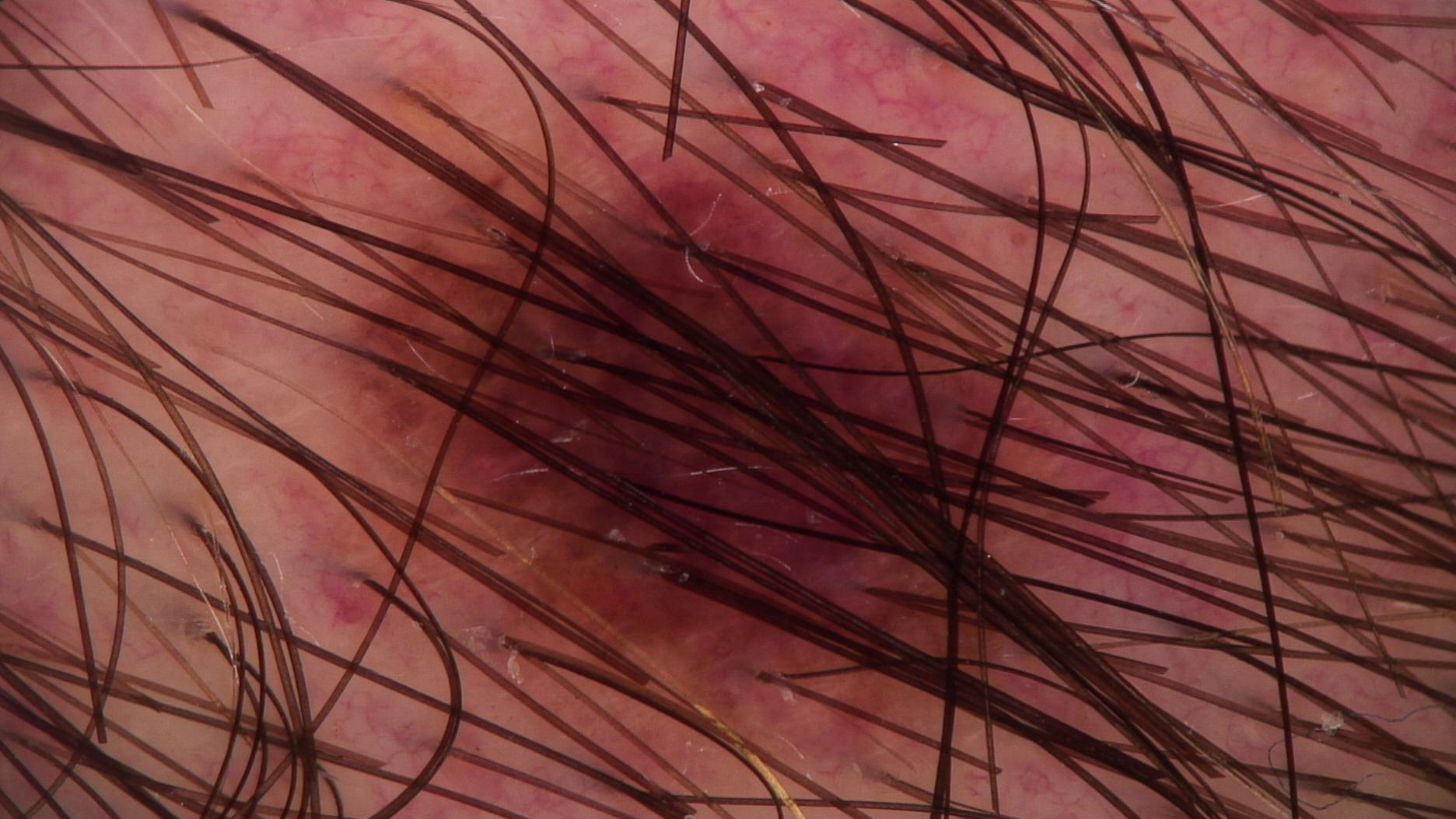}} {\includegraphics[width=.42\linewidth]{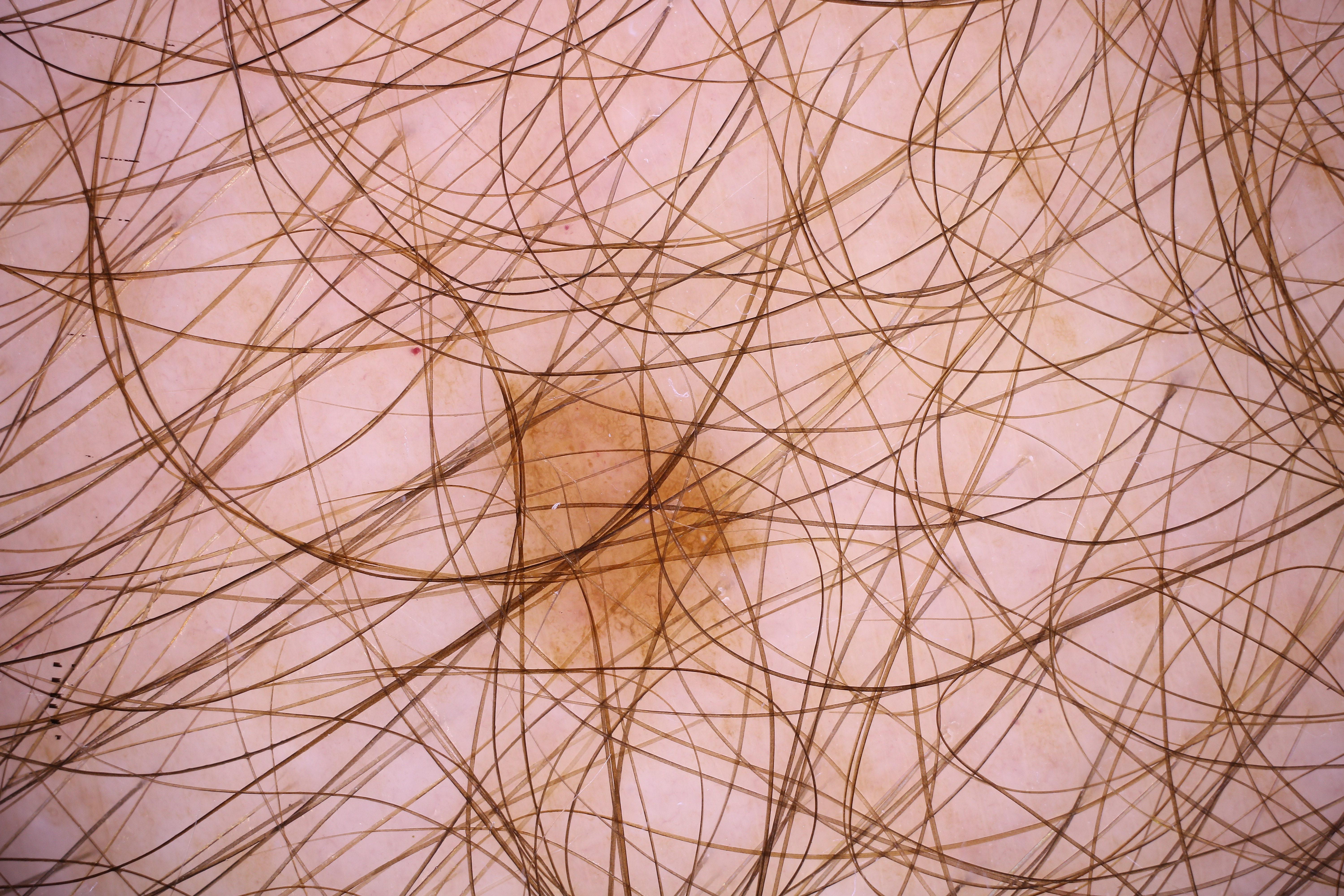}}}%
 \qquad
 \subfloat[\centering Images with a scale measurement in the picture]
 {{\includegraphics[width=.5\linewidth]{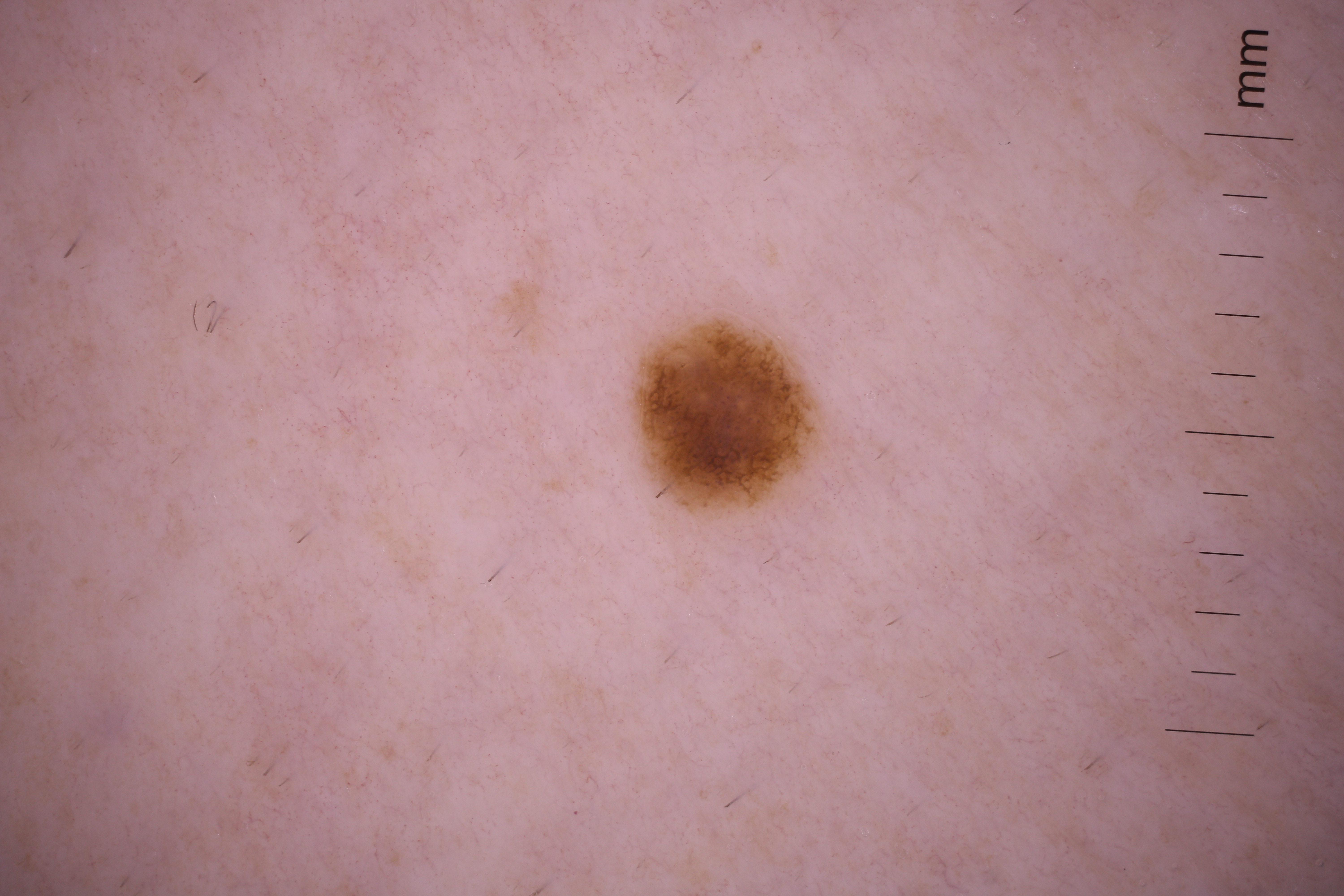}} {\includegraphics[width=.5\linewidth]{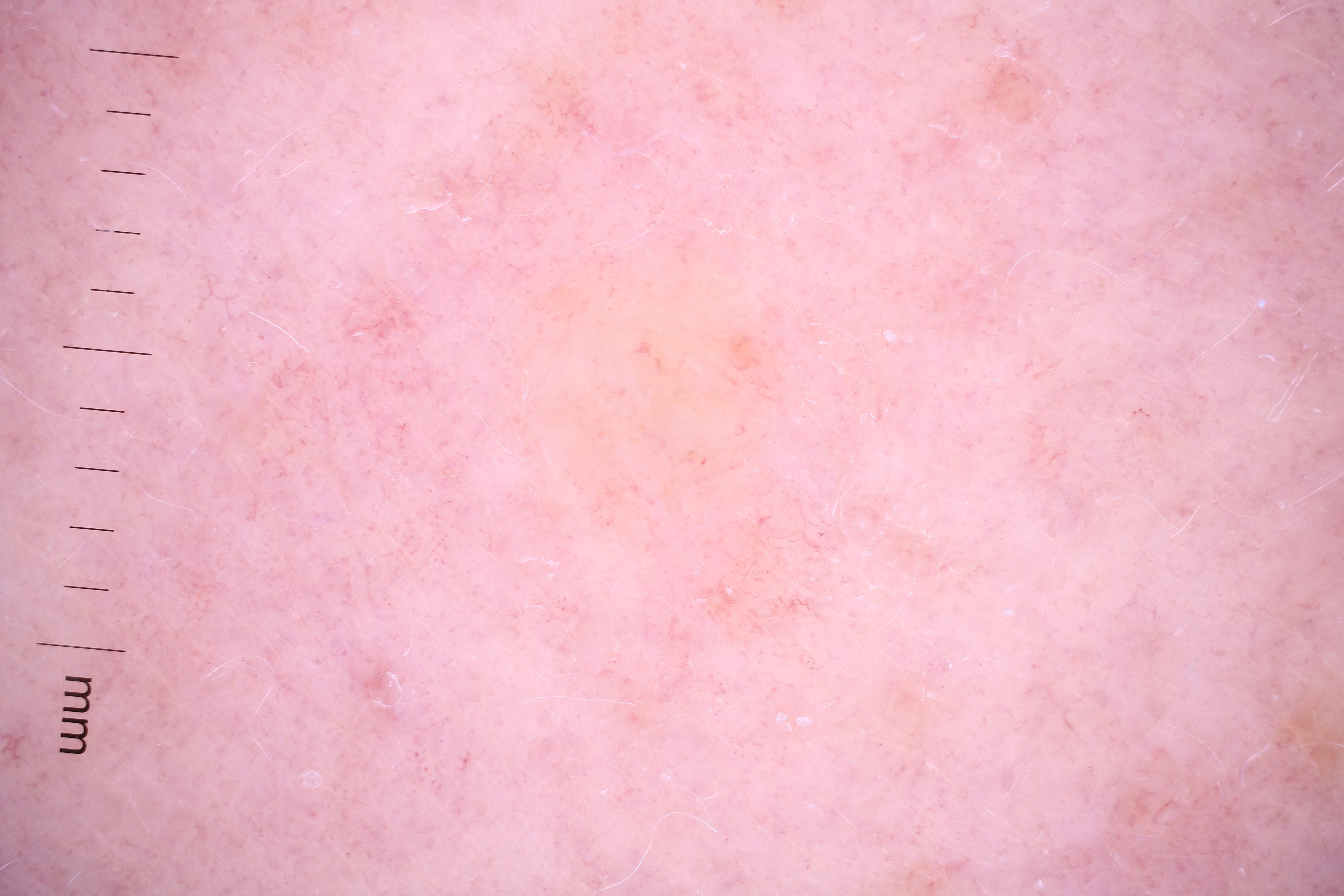}}}%
 \end{center}
 \caption{Sample images that include additional elements beyond the skin lesion.}
 \label{fig:sample-block-images}
\end{figure}

This study leverages the 2020 open-source SIIM-ISIC Challenge dataset, which comprises 33,126 dermoscopic images consisting of benign lesions (n=32,542 or 32,120 excluding duplicates) and malignant lesions (n=584 or 581 excluding duplicates)~\cite{rotemberg_patient-centric_2021}. Associated metadata for each image describes the patient's age, biological sex, the lesion's general anatomic site, a patient identifier, the ground truth classification label (malignant or benign), and the diagnosis if available (including melanoma, nevi, atypical melanocytic proliferation, caf\'e-au-lait macule, lentigo NOS, lentigo simplex, solar lentigo, lichenoid keratosis, seborrheic keratosis)~\cite{rotemberg_patient-centric_2021}. Notably, all images have been collected from 2,056 patients, facilitating holistic analysis of all lesions associated with a given patient (with 16 lesions associated with each patient, on average), while accounting for relative differences between lesions, which provides an essential source of context in clinical practice~\cite{rotemberg_patient-centric_2021}. 428 of these patients have at least one melanoma, while the remaining 1,628 patients have no melanoma lesions~\cite{rotemberg_patient-centric_2021}. Though the dataset contains images of skin lesions, some images also have hair in the image obscuring parts of the lesion and/or have a scale measurement present as seen in Figure~\ref{fig:sample-block-images}.

\subsection{Double-Condensing Attention Condenser Architecture Design}
The DC-AC architecture design was recently introduced as a self-attention neural network backbone which leverages DC-AC modules to achieve high accuracy and efficiency with a minimal computational footprint, supporting its use in TinyML applications~\cite{wongFasterAttentionWhat2022}. The architecture is shown below in Figure~\ref{fig:attendnext}. Each of the four computation branches has a series of convolutional layers and DC-AC blocks which enable them to learn distinct embeddings that are later merged.

A key feature of this architecture, which was designed using machine-driven exploration, is the use of DC-AC modules which enable selective self-attention by using both highly condensed feature embeddings, which are derived by condensing the original input features, and self-attention values~\cite{wongFasterAttentionWhat2022}. These modules build upon the previously introduced Attention Condensers, which are standalone, self-contained modules consisting of condenser layers, embedding layers, and expansion layers~\cite{wongTinySpeechAttentionCondensers2020}. The approach to condensing input features emphasizes activations close to other strong activations, resulting in an efficient selective attention mechanism~\cite{wongTinySpeechAttentionCondensers2020}. The DC-AC architecture design also employs a heterogeneous columnar design pattern which supports both independent and complex feature learning due to increasing columnar interactions as the level of abstraction increases~\cite{wongFasterAttentionWhat2022}. Stability and robustness are achieved by leveraging anti-aliased downsampling (AADS) throughout the network~\cite{wongFasterAttentionWhat2022}. 

This study uses the DC-AC architecture as described in~\cite{wongFasterAttentionWhat2022}.

\begin{figure}
  \centering
    \includegraphics[width=\textwidth]{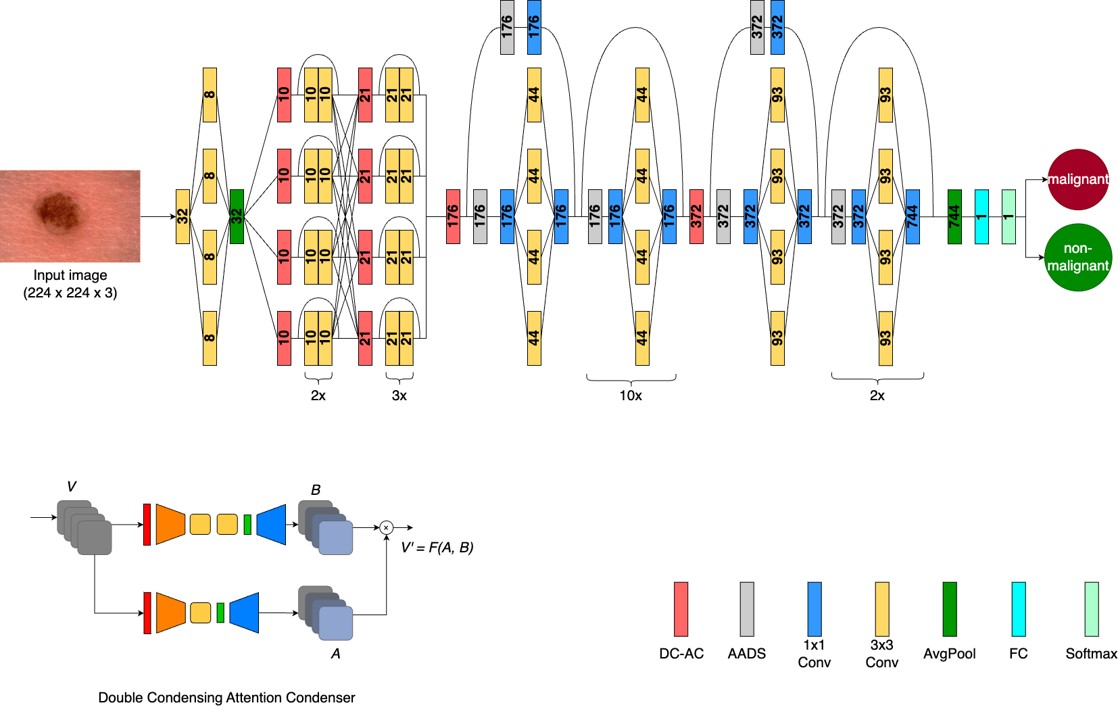}
  \caption{DC-AC architecture with condenser layers (orange), embedding layers (yellow) and expansion layers (blue) comprising DC-AC modules~\cite{wongFasterAttentionWhat2022}. The numbers each layer is annotated with correspond to the depth dimension of the layer.}
  \label{fig:attendnext}
\end{figure}

\subsection{Training Strategy}
In this study, transfer learning was employed by leveraging the DC-AC backbone pre-trained on ImageNet, while fine-tuning was performed using the ISIC 2020 dataset~\cite{rotemberg_patient-centric_2021}, to achieve an efficient network architecture design customized for dermoscopic image analysis and skin cancer detection with a low computational footprint. During training, the entire network was first frozen with the exception of the head. The head was trained on the ISIC 2020 dataset using the AdamW optimizer with 80 epochs, a learning rate of 0.00005, a weight decay of 0.01, and batch balancing. After the head was trained, the entire network was unfrozen and fine-tuned. Fine-tuning of the entire network was performed using the AdamW optimizer with 80 epochs, a learning rate of 0.000005, a weight decay of 0.01 and batch balancing. Cosine annealing was used to dynamically update the learning rates. 

Training was performed using a subset of the original dataset consisting of 22,860 images, including 437 malignant lesions. The remainder (9,841 images) were used for validation with 144 malignant lesions. 425 duplicate images were removed before the splitting of the dataset into training and validation. The images were divided by patients so that all images from the same patient were in the same set. Given that the training dataset consists of substantially more benign images than malignant images, batch balancing was performed using weighted random sampling to prevent bias towards the benign class. An additional 10,982 images with equal proportions of malignant and benign lesions comprise the test dataset and this unseen data was used to assess the trained architecture. 

Data augmentation was randomly applied to the training data including rotation (up to 90\textdegree), translation (up to 10\%), scale change (up to 20\%), shearing (up to 10\textdegree), colour jitter (up to 50\% brightness, contrast, saturation, and hue adjustment), horizontal flip, vertical flip, and random cropping and resizing (to 160x160 pixels). All validation images were resized to 160x160 pixels. 

\section{Results}

\begin{figure}
 \begin{center}
 \subfloat[\centering Malignant skin lesions]{{\includegraphics[width=.5\linewidth]{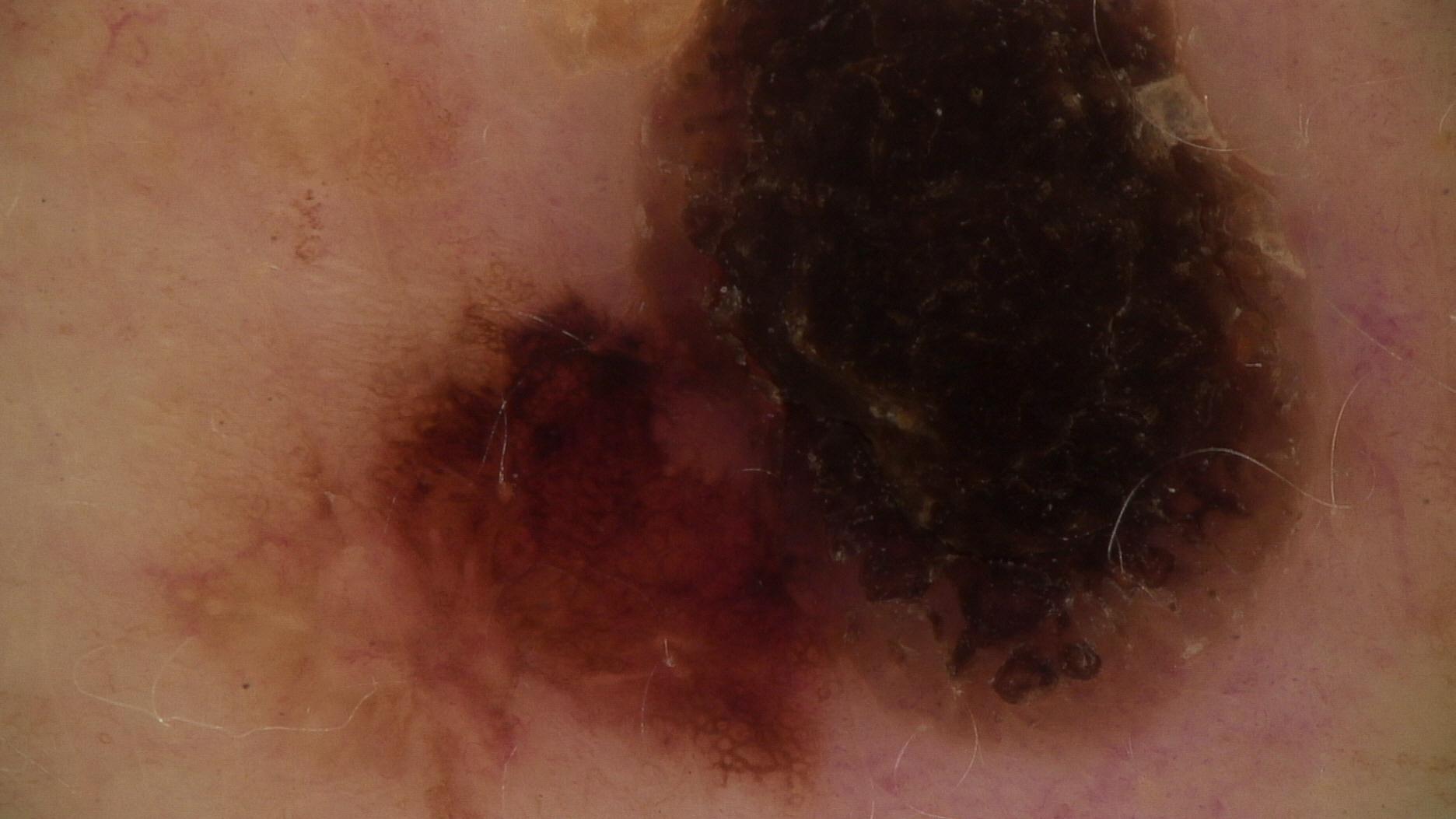}} {\includegraphics[width=.42\linewidth]{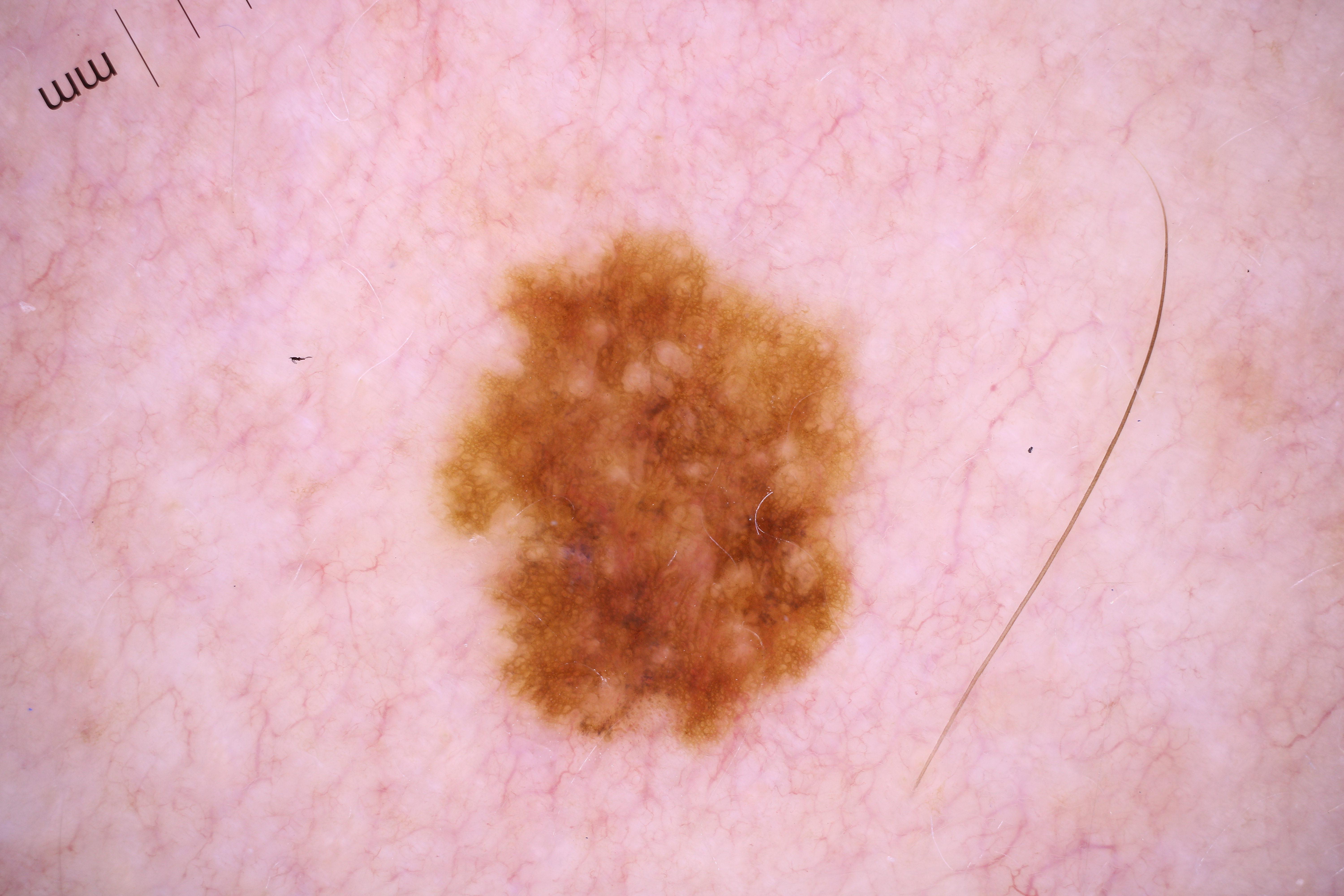}}}%
 \qquad
 \subfloat[\centering Benign skin lesions]
 {{\includegraphics[width=.5\linewidth]{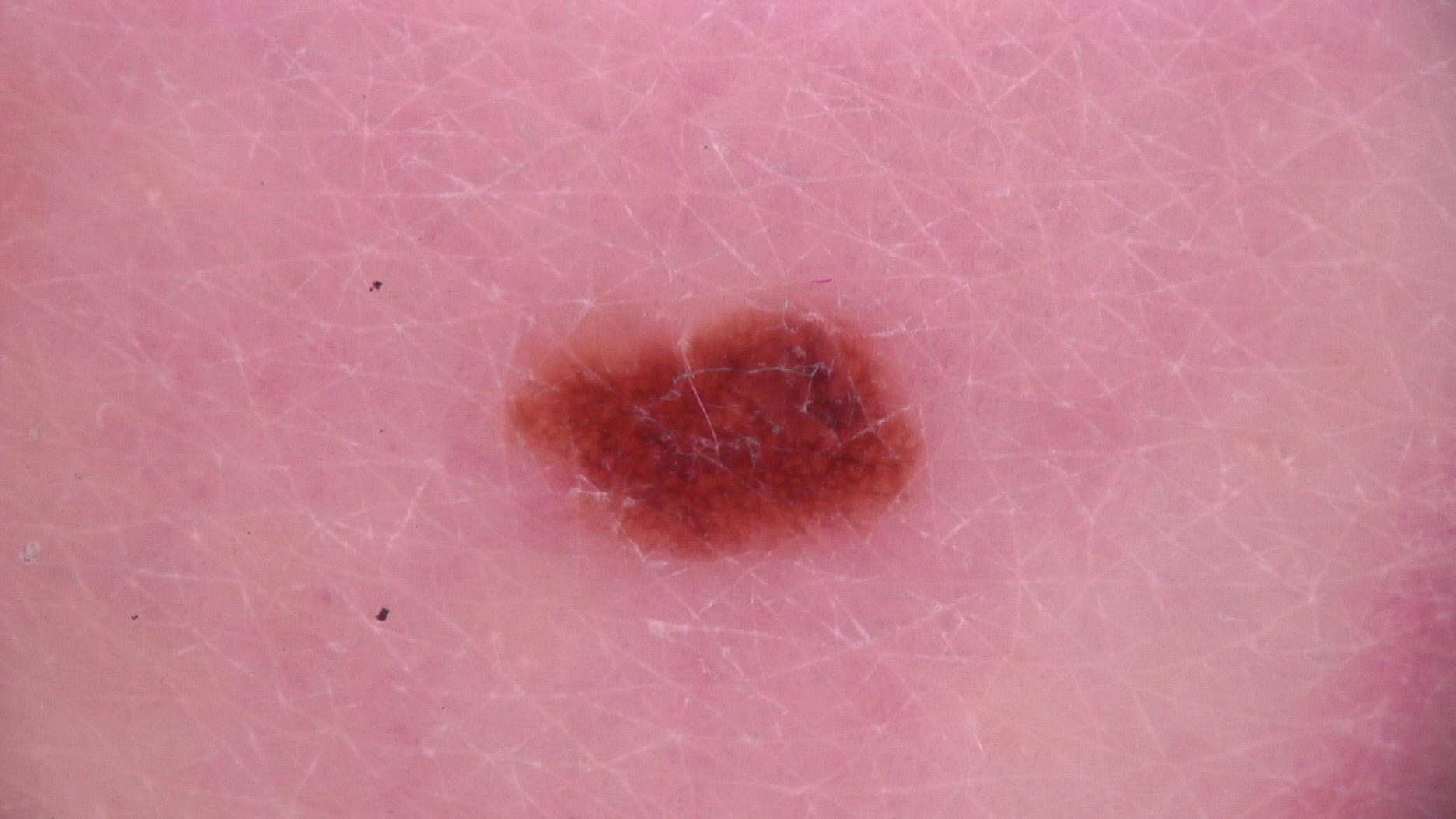}} {\includegraphics[width=.42\linewidth]{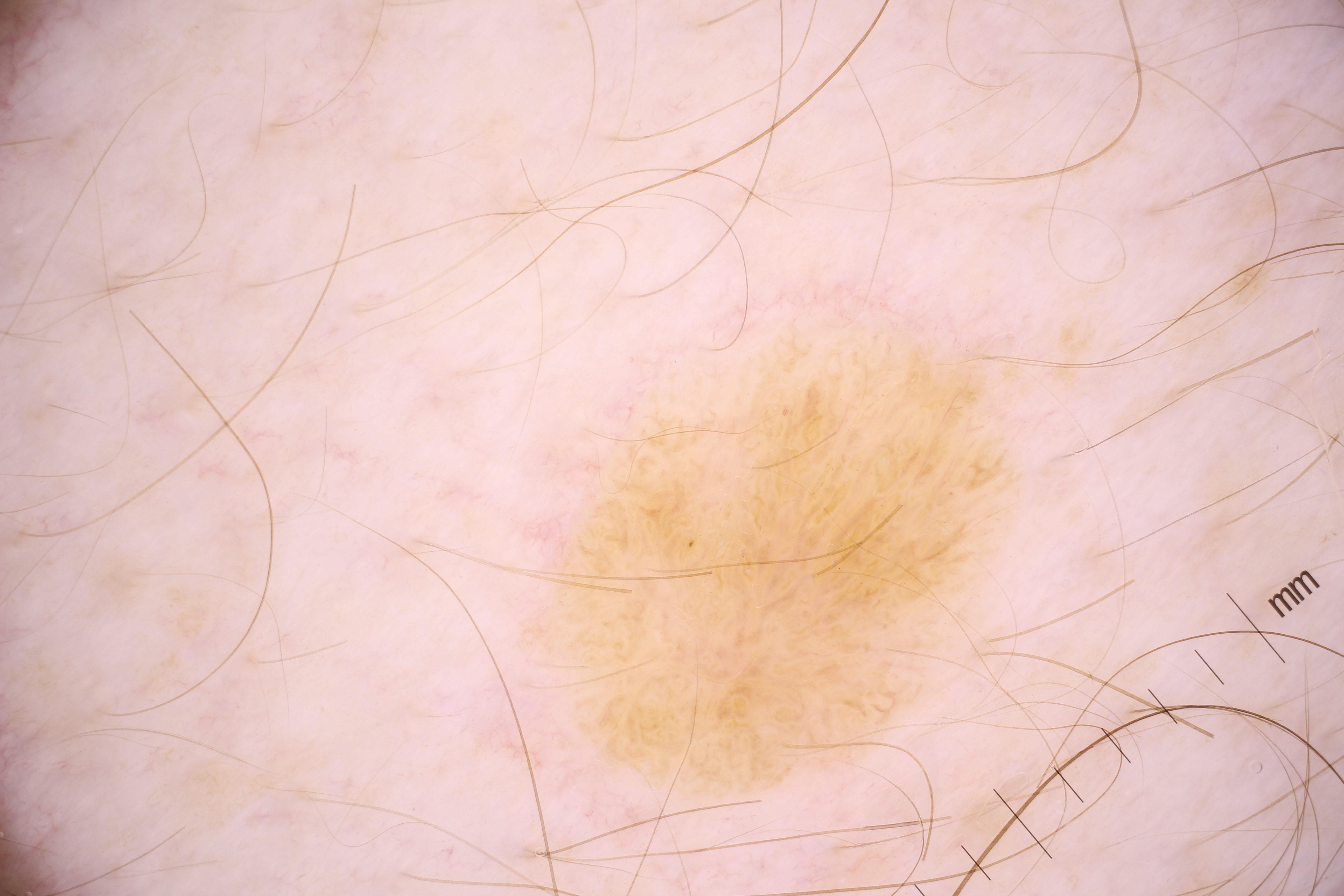}}}%
 \end{center}
 \caption{Sample skin lesions where the DC-AC design gave a correct prediction but the Cancer-Net SCa suite was incorrect.}
 \label{fig:sample-prediction-images}
\end{figure}

Using a train-val-test split, the efficacy of the proposed network architecture design was evaluated based on its performance on the original (non-augmented) data in each split. The SIIM-ISIC Kaggle competition platform~\cite{pan_siim-isic_2020} was leveraged to compute the performance on the unseen test set as the test labels were not publicly available anywhere else. Two scores are reported on the platform: public and private. The public score refers to the network's AUROC performance on 30\% of the test data with the private score calculating the AUROC performance on the remaining 70\% of the test data. On the training and validation split, the DC-AC achieves AUROC scores of 0.9787 and 0.8989 respectively. As seen in Table~\ref{test-results}, the test set performance for the DC-AC network architecture design outperforms the Cancer-Net SCa network architecture designs, achieving AUROC scores of at least 0.13 higher than the best Cancer-Net SCa architecture design. Sample skin lesions where the DC-AC design correctly predicted the class and the Cancer-Net SCa suite gave incorrect predictions can be found in Figure~\ref{fig:sample-prediction-images}.

\begin{table}
 \caption{Network architecture design performance on the ISIC 2020 test dataset.}
 \label{test-results}
 \centering
\begin{tabular}{ |c|c|c|c|c| } 
\hline
Network Architecture Design & Param. (M) & FLOPs (G) & Public Score & Private Score \\
\hline
\textbf{DC-AC} & \textbf{1.6} & \textbf{0.325} & \textbf{0.9045} & \textbf{0.8865} \\ 
 MobileViT-S~\cite{mehta2021mobilevit} & 5.6 & 2.03 & 0.8448 & 0.8566 \\
 Cancer-Net SCa-A~\cite{leeCancerNetSCaTailoredDeep2020} & 13.65 & 4.66 & 0.7538 & 0.7327 \\
 Cancer-Net SCa-B~\cite{leeCancerNetSCaTailoredDeep2020} & 0.80 & 0.43 & 0.7697 & 0.743 \\
 Cancer-Net SCa-C~\cite{leeCancerNetSCaTailoredDeep2020} & 1.19 & 0.40 & 0.737 & 0.7333 \\
 \hline
\end{tabular}
\end{table}

\section{Discussion}

The test set performance for the self-attention model via attention condensers also outperforms the self-attention model via transformers, MobileViT-S~\cite{mehta2021mobilevit}. MobileViT-S is a state-of-the-art vision transformer that is designed to provide state-of-the-art performance balance between computational, architecture complexity, and accuracy. However, in terms of parameters, DC-AC has significantly less parameters than the next best network architecture design (1.6 M vs 5.6 M). The DC-AC design also has the least number of FLOPs (G) or in other words, the lowest number of floating point operations needed for a single forward pass. As such, the DC-AC design has better computational performance compared to the other models.

\section{Conclusions}
In this paper, we investigated leveraging the efficient self-attention structure of a DC-AC architecture to detect skin cancer in skin lesion images. Evaluation using the previously unseen ISIC 2020 test set showed that the proposed DC-AC architecture can increase skin cancer detection performance compared to previously released non-ensemble network architectures. Specifically, using the DC-AC design led to a public AUROC score of 0.9045 and a private AUROC score of 0.8865, which is over 0.13 above the best Cancer-Net SCa network architecture design. Subsequently, future work involves iterating on the design of the selected network architecture and refining the approach to generalize to other forms of cancer.

\begin{ack}
The authors thank Hayden Gunraj for his invaluable support, exemplifying both his intelligence and personable nature, which greatly enriched the project's success.
\end{ack}

{
\small

\bibliography{neurips_2024}
}

\end{document}